\documentclass[12pt]{article}

\usepackage{graphicx}

\newcommand{\beq}{\begin{equation}}
\newcommand{\eeq}{\end{equation}}
\newcommand{\beqa}{\begin{eqnarray}}
\newcommand{\eeqa}{\end{eqnarray}}
\newcommand{\beqar}{\begin{eqnarray*}}
\newcommand{\eeqar}{\end{eqnarray*}}

\begin{document}
\thispagestyle{empty}

\hfill{\sc UG-FT-226/08}

\vspace*{-2mm}
\hfill{\sc CAFPE-96/08}

\vspace{32pt}
\begin{center}

\textbf{\Large Hawking evaporation of cosmogenic black holes\\
in TeV-gravity models}
\vspace{40pt}

Petros Draggiotis$^{1}$, Manuel Masip$^{1}$, 
Iacopo Mastromatteo$^{1,2}$
\vspace{12pt}

\textit{
$^{1}$CAFPE and Departamento de F{\'\i}sica Te\'orica y del
Cosmos}\\ 
\textit{Universidad de Granada, E-18071 Granada, Spain}\\
\vspace{8pt}
\textit{$^{2}$Dipartimento di Fisica Teorica}\\ 
\textit{Universita degli Studi di Trieste, I-34014 Trieste, Italy}\\
\vspace{16pt}
\texttt{pdrangiotis@ugr.es, masip@ugr.es, iacopomas@infis.univ.trieste.it}
\end{center}

\vspace{40pt}

\date{\today}

\begin{abstract}

We study the properties of black holes of mass $10^4$--$10^{11}$ GeV 
in models with the fundamental scale of gravity at the TeV. These 
black holes could be produced in the collision of a ultrahigh energy 
cosmic ray with a dark matter particle in our galactic halo or with 
another cosmic ray. We show that QCD bremsstrahlung and pair production 
processes are unable to thermalize the particles exiting the black
hole, so a {\it chromosphere} is never formed during Hawking 
evaporation. We evaluate with HERWIG the spectrum of stable 
4-dim particles emitted during the Schwarzschild phase and find that 
in all cases it is peaked at energies around 0.2 GeV, with an 
approximate 43\% of neutrinos, 28\% of photons, 16\% of electrons 
and 13\% of protons. Bulk gravitons are peaked at higher energies, 
they account for a 0.4\% of the particles (16\% of the total energy) 
emitted by the most massive black holes in $n=6$ extra dimensions 
or just the 0.02\% of the particles (1.4\% of the energy) emitted 
by a 10 TeV black hole for $n=2$.

\end{abstract}


\newpage

\section{Introduction}

The coexistence of the electroweak and the Planck scales
has been the main motivation for model building in 
particle physics during the past 30 years. Early proposals like
technicolor or supersymmetry could explain dynamically the
hierarchy between these two scales, although the new physics that
they suggest has been so far absent in collider experiments.
More recent proposals 
offer new and very interesting possibilities. In particular, 
the presence of extra dimensions could imply a 
fundamental scale of gravity $M_D$ much lower than 
$M_P=G_N^{-1/2}\approx 10^{19}$ GeV 
\cite{ArkaniHamed:1998rs}. 
If $M_D$ were near the TeV region, 
then the electroweak scale would just introduce a 
{\it little hierarchy} problem, which could be 
easier to solve consistently with collider data \cite{Yao:2006px}. 

In contrast with the usual scenario, 
an amusing feature in these TeV-gravity models is that $M_D$ 
is at accessible energies and the {\it transplanckian} regime 
can in principle be probed 
\cite{Banks:1999gd}.
Actually, due to the spin 2 of
the massless graviton the collision of two point-like particles 
at transplankian energies ($\sqrt{s}\gg M_D$) and 
large distances ($b\gg M_D^{-1}$) should be dominated by gravity.
In such collision one would expect the collapse of the 
two particles into a black hole (BH) of 
mass $M\approx \sqrt{s}$ with an approximate cross section 
\beq
\sigma=\pi r_{H}^2\;,
\label{eq1}
\eeq
where $r_{H}$ is the radius of the BH horizon. Note also that 
the exchange of the large momenta required to {\it see} 
string resonances \cite{Cullen:2000ef} or other quantum gravity 
effects would take place here at shorter distances: 
all the details of the complete theory that 
describes quantum gravity will be {\it screened} by the BH horizon.
As the collision energy increases, $r_{H}$ grows and 
the collapse into a BH involves larger distances, a regime 
where classical gravity (strongly coupled but with no loops
\cite{Giudice:2001ce}) should work well.

The production of microscopic BHs at the LHC has been extensively
considered in the literature 
\cite{Giudice:2001ce,Cheung:2002aq,Harris:2003db,
Winstanley:2007hj,Meade:2007sz}. 
It seems difficult, however, to imagine
a framework able to accommodate such expectations. First of
all, the mechanism that defines a consistent theory of gravity
should also manifest below $M_D$ while being consistent with
all precision data. If, for example, gravity derives 
from string theory, the string scale $M_S$ will be a loop factor 
smaller than $M_D$ \cite{Cullen:2000ef}:
\beq
M_S^{n+2}={g^4\over 8 \pi} M_D^{n+2}\;,
\label{eq2}
\eeq
with $n$ the number of large extra 
dimensions\footnote{See the appendix in \cite{Cavaglia:2002si} 
for the relation of $M_D$ with $M_*$}. Second, 
even if $M_D$ is as low as a few TeV, the $\sqrt{s}$ at
the parton level accessible at the LHC can not be 
much larger. The BHs that one may expect there will be 
near the threshold $M_D$, where 
they would appear almost indistinguishable from a massive 
string mode.

In this paper we focus on much {\it larger} mini BHs, of mass 
up to $10^{11}$ GeV. They could result from the 
collision of a cosmic ray (a proton or a cosmogenic neutrino)
with a dark matter particle in our galactic halo 
or with another cosmic ray. First we estimate the 
production rate of these {\it cosmogenic} BHs.
Then we describe their Hawking evaporation (temperature, emision 
rate, lifetime) including in the evaluation the 
greybody factors for the different particle species. 
An important point that we address is the formation of a 
chromosphere or a photosphere around the BH during its 
evaporation. We show that, despite the large temperature  
(between 1 and 300 GeV), strong or electromagnetic interactions 
are unable to thermalize the particles exiting the BH, so their
energy is not altered by this factor. 
Finally, using the code HERWIG \cite{Corcella:2000bw}
(which simulates the fragmentation
of colored particles as well as particle decay) we 
derive the spectrum of stable species 
(protons, electrons, photons and neutrinos) and of
bulk gravitons 
produced in the evaporation of a BH of given mass. 
Previous works on BH production by cosmic rays refer
to the collision of a cosmogenic neutrino with an
atmospheric nucleon \cite{Feng:2001ib} or a cosmic
ray with a nucleon in the interestellar medium
\cite{Barrau:2005zb}.

\section{Black hole production at ultrahigh energies}

We observe a flux of cosmic rays (most of them 
are protons free or bound in nuclei) \cite{Yao:2006px}
that reach the earth with energies of 
up to $10^{11}$ GeV. Their production and propagation 
induces a flux of (still unobserved) cosmogenic 
neutrinos peaked at energies near $10^{9}$ GeV 
\cite{Semikoz:2003wv}. For example, a km$^2$ area in the upper 
limit of the atmosphere would receive around $10^4$ 
protons and $10^3$ neutrinos\footnote{We take the  
neutrino flux in Fig.~2 of Ref.~\cite{Semikoz:2003wv}.} 
of energy between $10^8$ and $10^{11}$ GeV per year.

In addition, it is thought that 90\% of the matter in our
galaxy ($10^{12}$ solar masses) is {\it dark}: $10^{69}$ GeV
of mass in a sphere of 200 kpc, with an 
approximate density profile\footnote{We assume a CUSP dark 
matter distribution \cite{Navarro:1995iw}.}:
\beq
\rho(r)\approx {\rho_0\over {\left( {r\over R}\right) 
\left( 1+ {r\over R}\right)^2}}\;,
\label{eq3}
\eeq
with $R=20$ kpc (we are at 8 kpc from the center, 
$1\; {\rm kpc}=3\times 10^{19}$ m).
This dark matter would be constituted by a weakly interacting
massive particle $\chi$ of mass $m_\chi\approx 100$ GeV, 
although $m_\chi$ could go from 10 MeV to 10 TeV if its 
interaction strength goes from gravitational to 
strong \cite{Feng:2008ya}.

Therefore, we find two types of elementary processes involving 
center of mass energies above $M_D$. 

\noindent {\it (i)} The collision of a cosmic ray of energy 
$E$ with a 
dark matter particle. Here $\sqrt{s}=\sqrt{2 m_\chi E}$ may go 
up to $10^7$ GeV. The
number of interactions to produce a BH 
per unit time and volume in terms 
of the dark matter density ($\rho_\chi=\rho/m_\chi$), 
the flux (integrated over 
all solid angles), and the cross section is
\beq
{{\rm d}^2N\over {\rm d}t\; {\rm d} V}= 
4 \pi \int {\rm d}E \;\sigma_{i \chi} (s)\;
\frac{{\rm d}\phi_i}{{\rm d}E}\; \rho_\chi \;.
\label{eq4}
\eeq

\noindent {\it (ii)} 
The collision of two cosmic rays. The center of mass energy 
for a relative direction $\theta$ between the two particles is
$\sqrt{s}=\sqrt{2E_1 E_2(1-\cos\theta)}$, 
which can reach $10^{12}$ GeV.
Given the two fluxes and the cross section, the number of
interactions producing BHs is just
\beq
{{\rm d}^2N\over {\rm d}t\; {\rm d} V}= 16 \pi^2 
\int {\rm d}E_1\; {\rm d}E_2\; {\rm d}\cos\theta
\;\sigma_{ij} (s)\;\sin\theta/2\;
\frac{{\rm d}\phi_i}{{\rm d}E_1} \;
\frac{{\rm d}\phi_j}{{\rm d}E_2}\;.
\label{eq5}
\eeq

\begin{figure}
\begin{center}
\includegraphics[width=0.5\linewidth]{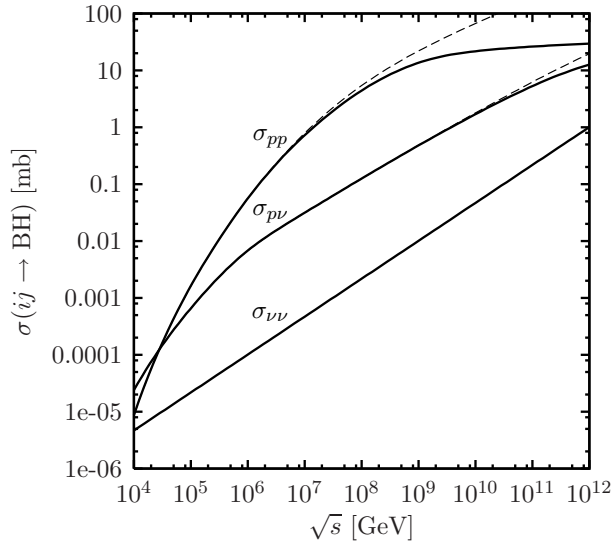}
\caption{
Cross sections to produce a BH for $n=2$ and $M_D=1$ TeV. 
\label{fig1}}
\end{center}
\end{figure}

It is then necessary to evaluate the BH production 
cross section $\sigma_{ij}$,  
where $i,j$ label a proton, a neutrino or a $\chi$. 
We will asume that all particles except for the graviton
live on a 4 dimensional brane, that there are $n$ flat 
extra dimensions of common length, and that $r_{H}$ is
just the higher dimensional Schwarzschild radius,
\beq
r_{H}=\left({2^n\pi^{n-3\over 2}\Gamma\left({n+3\over 2}\right)
\over n+2}\right)^{1\over n+1}
\left({M\over M_D}\right)^{1\over n+1}
     {1\over M_D}\;,
\label{eqrbh}
\eeq
where $M\approx \sqrt{s}$ is the BH mass. 
It is then easy to find the cross section 
$\sigma_{\nu\nu}$ (or $\sigma_{\nu\chi}$)
between two pointlike particles; we plot it in 
Fig.~\ref{fig1} for $n=2$ and $M_D=1$ TeV. 

The calculation is not that simple in $p\nu$ or $pp$  
collisions, as at short distances the interaction involves
partons. Actually, the expression for pointlike particles 
still holds at very large energies, when $r_{H}>1$ fm and
the BH is large enough to contain the proton. 
In that case the neutrino interacts coherently with
the whole proton and $\sigma_{p\nu}(s)$  
just coincides with $\sigma_{\nu\nu}(s)$.
At low $s$, the case extensively
discussed in previous literature, the neutrino interacts with
a single parton carrying a fraction $x$ of the proton 
momentum and they collapse into a BH of mass $M=\sqrt{x s}$.
The cross section is
\beq
\sigma_{p\nu}(s)=\int_{M_D^2/s}^1 {\rm d}x 
\left(\sum_i f_i(x,\mu) \right) \sigma(xs)\;. 
\label{eq7}
\eeq
At higher energies more partons in the low $x$ 
region are able to produce BHs, $\sigma_{p\nu}$ 
quickly grows and Eq.~\ref{eq7} becomes no longer 
reliable. A value of $\sigma_{p\nu}\approx 20$ mb  
indicates that {\it all} neutrinos approaching with 
impact parameter smaller than the proton radius
will interact with a parton to form a BH. 
Obviously, increasing the energy in the collision we can
not get larger cross sections unless the typical BH produced
has a $r_{H}$ similar to the proton radius (the 
regime discussed before).

Therefore, we distinguish three regimes in  
the $p\nu$ interaction to produce a BH. 
At low energies the 
neutrino interacts with a single parton, 
the cross section grows with $s$ but the average mass 
$M$ of the BH is roughly constant and close to the 
threshold $M_D$. Once it approaches 20 mb, 
the cross section remains constant, but the 
typical BH mass increases with the energy. In this 
regime the process involves multiple scattering, 
in the sense that the BH produced in the collision 
will also trap {\it spectator} partons. Finally, at 
higher energies (not in the plot) the proton becomes
pointlike and the cross section grows just like 
$\sigma_{\nu\nu}$.
In Fig.~\ref{fig1} we have modelled a smooth transition between 
these regimes by {\it discretizing} the proton and 
discounting the overlapping between parton cross sections.
The $pp$ collision (also in Fig.~1) is completely
analogous, although the $20$ mb bound on $\sigma_{pp}$ is
saturated at $\sqrt{s} \sim 10^7$ GeV (notice that this
effect is still negligible at LHC energies).

\begin{figure}
\begin{center}
\includegraphics[width=0.5\linewidth]{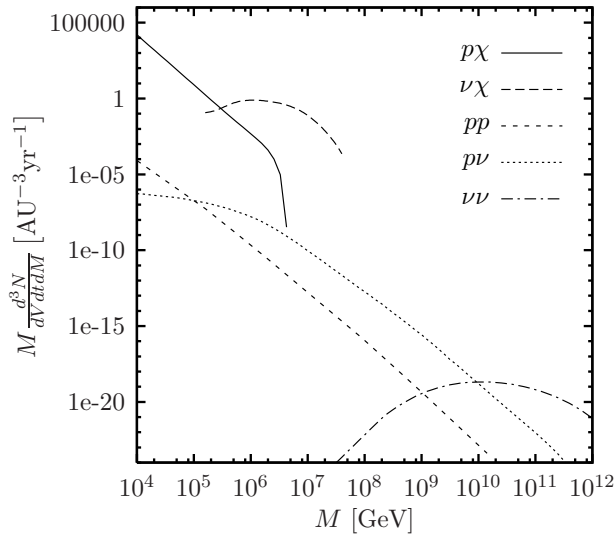}
\caption{
Spectrum of BHs produced in cosmic ray collisions
for $n=2$, $M_D=1$ TeV, $m_\chi=100$ GeV, and the
upper neutrino flux in \cite{Semikoz:2003wv}.
\label{fig2}}
\end{center}
\end{figure}

We can now estimate the BH production rate in the
two types of processes discussed above. In 
Fig.~\ref{fig2} we
plot the number and mass distribution of BHs produced 
in the collision of a cosmic ray (a proton or a cosmogenic 
neutrino) with another cosmic ray or with a dark matter 
particle per year and per cubic 
astronomical unit (1 AU = 1.5 $\times 10^{11}$ m, the mean
earth-sun distance). 
We have taken the expected dark matter density 
at our position in the galaxy ($\rho=0.3$ GeV/cm$^3$) and
$m_\chi=100$ GeV. 

\section{Black hole evaporation}

Once produced 
the BH will go through a fast {\it balding phase}, where it 
loses its {\it gauge hair} and asymmetries, and a relatively
brief {\it spin-down phase} \cite{Winstanley:2007hj}. 
Most of its lifetime the BH will be Schwarzschild-like, 
and it will emit Hawking radiation \cite{Hawking:1974rv} 
with an approximate black body spectrum of temperature 
\beq
T={n+1\over 4\pi r_{H}}\;.
\label{eqT}
\eeq
Notice that given 
$r_{H}$, the energy radiated by the BH will not depend 
(up to factors of order one) on the number of extra dimensions:
on dimensional grounds 
$\dot{E}\sim A_{2+n} T^{4+n}\sim 1/r_{H}^2\sim T^2$. Since a 
bulk and a brane field {\it see} a BH of the same temperature
($T$ is constant along the BH surface), the BH will emit a similar
amount of energy of both fields \cite{Emparan:2000rs}. 

The spectrum of particles exiting the BH must cross 
the strong gravitational potential near the horizon in 
order to escape to infinity \cite{Page:1976df}. This effect 
is usually described
in terms of the {\it greybody factors} 
$\Gamma_s=\sigma_s/A_{2+n}$, where $\sigma_s$ is the absortion
cross section for a particle of spin $s$ living in 
4+$n$ dimensions and $A_{2+n}$ is the BH area seen by that 
particle. The average emission rate for a 4-dim 
particle of energy $\omega$ is then
\beq
{{\rm d}^2 N_i\over {\rm d}\omega\;{\rm d}t}=
{A_2\over 8\pi^2}{c_i\Gamma_s\; \omega^2\over e^{\omega/T}
-(-1)^{2s}}\;,
\label{N}
\eeq
being $c_i$ the multiplicity of the species. Throughout the 
paper we use for the 4-dim particles the (numerical) greybody
factors given in \cite{Harris:2003eg}, together with the expressions in 
\cite{Cardoso:2005vb} for the higher dimensional graviton. 
The emission 
into bulk gravitons can be significant for a large number 
of extra dimensions, for example, it accounts
for a 16\% (1.6\%) of the radiated energy for $T=10$ GeV and 
$n=6$ ($n=2$).
In our numerical estimates we will only consider the emission
of (relativistic) 
particles lighter than the BH temperature, including at 
$T<1$ GeV no colored particles but pions as fundamental degrees 
of freedom. Notice that the emission will be dominated 
by quarks and gluons, as these cosmogenic BHs have a temperature
above $\Lambda_{QCD}$.

\begin{figure}
\begin{center}
\includegraphics[width=0.5\linewidth]{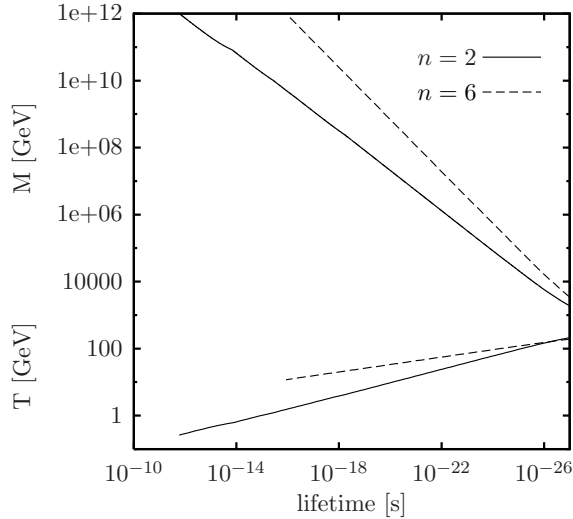}
\caption{
Correlation between mass, temperature,
and lifetime of a BH for $M_D=1$ TeV and $n=2,6$. 
\label{fig3}}
\end{center}
\end{figure}

It is straightforward to integrate (\ref{N}) over all frequencies
and sum over all particle species to deduce the BH mass loss 
per unit time. In Fig.~\ref{fig3} we plot the correlation between 
lifetime, mass and temperature for $M_D=1$ TeV and $n=2,6$. 
We see, for example, that a $10^{11}$ GeV BH lives (at rest) 
around $10^{-14}$ s and has an initial temperature of 0.6 GeV
for $M_D=1$ TeV and $n=2$.

\section{Chromosphere around evaporating Black Holes}

An important point raised by Heckler in 1996 \cite{Heckler:1995qq}
is that BHs above some
critical temperature $T_c\approx m_e/\alpha^{5/2}$  may form a 
surrounding photospere (a plasma of electrons and photons) of
outer temperature $T\approx m_e$.
The photosphere would thermalize to this low temperature
the particles exiting the BH, changing dramatically
the initial greybody spectrum.
Moreover, he argued that QCD processes would also define 
a chromosphere in
BHs of $T> \Lambda_{QCD}$, a fact that could affect, for example,
the Page-Hawking limits on primordial BHs.
After Heckler's initial claim 
there have been several analysis of photo/chromosphere formation
\cite{Cline:1998xk,Anchordoqui:2002cp},
although its existence has been considered controversial. 
In particular, none of the several codes simulating BH production 
at the LHC \cite{Harris:2003db} has included its effect.

Recently two different groups 
(Alig, Drees and Oda \cite{Alig:2006up}, and 
Carr, MacGibbon and Page \cite{MacGibbon:2007yq}) 
have reanalyzed this issue and have concluded that 
bremsstrahlung and pair production
processes are not able to form a photo/chromosphere 
around evaporating BHs. The key argument
is that the scattering of two particles radiated away from a BH 
can not be treated in the same way as the ordinary
collision of a particle against a target, since the
kinematics are completely different. In the radial 
case the particles are not coming from an infinitely far past, 
they are created in a definite point of space-time. In
addition, the two particles are {\it always} separating
(never approaching), a fact that introduces a maximum radius in
which the process can take place. Although a general formalism
to describe the radial scattering is not available, it is clear
that the calculation of the interaction rate as
\beq
\Gamma = \langle \sigma v \rho \rangle
\label{gamma}
\eeq
can lead to incorrect results. In this section 
we use the approach in \cite{MacGibbon:2007yq}
to show that the higher dimensional
BHs of mass up to $10^{11}$ GeV 
under analysis here do not form a chromosphere: quarks and
gluons escape the BH (at distances around $1/\Lambda_{QCD}$ they
fragment into hadrons, see next section) with basically the 
initial energy. This result is consistent with the detailed
MonteCarlo simulation in \cite{Alig:2006up}.

To be definite we will take a very hot (LHC-like) BH, 
with a temperature around $T=100$ GeV, 
and $M_D=1$ TeV for $n=3$. We find that such BH emits 
quarks or antiquarks of energy 
$E\approx 3\;T$ with a frequency of one per  
$\tau \approx 0.8/E = 0.8/\gamma m_q$ (see Table 1).
Once a quark exits the BH, 
it will be localized along an approximate radial distance of
$\lambda\approx 1/E$ (its reduced Compton wavelength).
Therefore, it can {\it overlap} significantly only with two
or three other quarks, being the rest of quarks 
separated by a distance of order $1/E$ or larger. 

\begin{table}[htdp]
\begin{center}
\begin{tabular}{|c|c|c|}
\hline
$n$ & $1/\nu_e$ & $1/\nu_q$ \\
\hline
2 & $16$ & $1.3$ \\
3 & $9$ & $0.8$ \\
4 & $7$ & $0.7$ \\
5 & $5$ & $0.4$ \\
6 & $4$ & $0.3$ \\
\hline
0 & $175$ & $20$ \\
\hline
\end{tabular}
\caption{Average distance between consecutive
electrons ($1/\nu_e$) and quarks ($1/\nu_q$) in
reduced Compton wavelength units ($1/E$). The last 
line refers to the primordial 4-dimensional BHs 
in \cite{MacGibbon:2007yq}}
\end{center}
\label{Suppression Parameters}
\end{table}

It is easy to see that the probability that the quark interacts 
with one of these non-overlapping quarks is negligible
just because they are {\it too far}.
The argument goes as follows.
Suppose that the quark $q_1$ exiting the BH interacts with
a quark $q_0$ at a distance of $k>1$ Compton wavelengths 
($1/E$) in the BH frame. 
The bremsstrahlung process is best understood in the 
rest frame of $q_0$; there $q_1$ interacts with the field 
generated by the static $q_0$, goes off-shell 
and emits a gluon. 
Since the world line of $q_1$ has a {\it beginning},  
its shorter distance $d$ with $q_0$ will correspond to the moment
when it appears in the BH horizon (except if it is emitted within
a small solid angle of order $1/\gamma$, see below). 
Suppose that right in that moment $q_1$
receives a gluon previously emited by $q_0$; it is 
straightforward to find that this gluon has been traveling a 
time/distance\footnote{The contributions out 
of the light cone are exponentially suppressed in the propagator.}
\beq
t = d \approx k\;\gamma/E=k/m_q\;
\label{D}
\eeq
in the $q_0$ frame. 
Since $d\approx k/m_q\sim k/\Lambda_{QCD}$ is 
the minimal distance with $q_1$ that $q_0$ may detect, 
the interaction will take place at typical distances 
where QCD is not effective (or,
equivalently, through the exchange of momenta below the 
infrared cutoff $\Lambda_{QCD}$). In \cite{MacGibbon:2007yq} 
the suppression in this radial cross section 
with minimal distance $d$ is 
estimated by the contribution of impact paramenters 
$b\ge d$ in a regular cross section (see Fig.~4). 
This suppression for the interaction
between the two quarks can be also
understood in the BH or the center of mass frames. 
There the time/distance 
that the gluon has been traveling is much shorter, 
$t'\approx k/E$. However, this time is already too large for the 
virtual gluon of energy of order $E$ required by the bremsstrahlung 
kinematics. Notice that it is the energy of this virtual
gluon, and not its invariant {\it off-shellness} 
$Q^2 \ge \Lambda_{QCD}^2$, what determines the maximum time
that it can travel without violating Heisenberg uncertainty 
principle. 
\begin{figure}
\begin{center}
\includegraphics[width=0.4\linewidth]{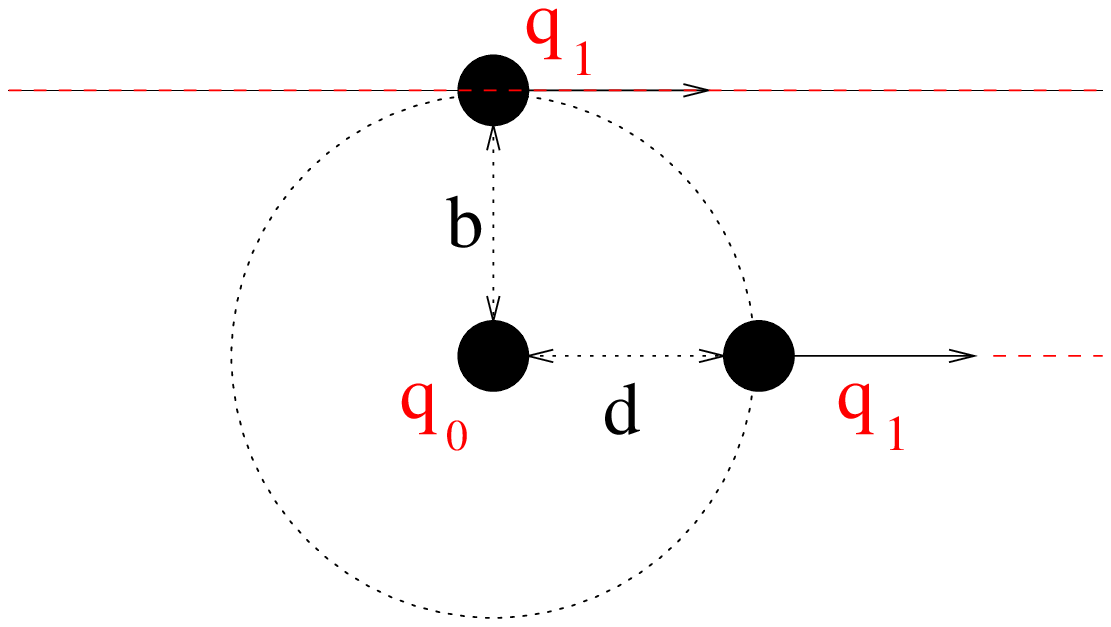}
\caption{
In a regular scattering $q_1$ comes from $-\infty$
with impact parameter $b$, whereas here $q_1$ is
created at a minimal distance $d$.
\label{fig4}}
\end{center}
\end{figure}

The {\it causality} or minimal distance argument just 
outlined suffices to 
disregard the formation of a photospere (or a chromosphere) in
high temperature 4-dim BHs. The reason is that the
average distance $(\nu_e E)^{-1}$ 
(or $(\nu_q E)^{-1}$) between consecutive electrons 
(quarks) emitted by the BH is 175 (20) times larger than the 
typical distances dominating the QED (QCD) bremsstrahlung 
cross section. Moreover, this distance (in reduced 
Compton wavelength units) does not depend on
the BH temperature, as both the number of particles and
their typical energy grow linearly with $T$.
In $4+n$ dimensions we find (see Table 1) that
this argument always suppresses the formation of a 
photosphere, but not of a chromosphere if $n>2$. In these cases
the distance between a quark and two or three 
other quarks is small (they overlap). We will
then use a second argument 
\cite{Alig:2006up,MacGibbon:2007yq,Klein:1998du} 
that disfavors the multiple 
interactions required to form a chromosphere: 
the existence of a maximum radius, 
$r_{brem}\approx 1/\Lambda_{QCD}$ in the
BH frame, where the interaction can take place. 

This maximum radius appears in the interaction of 
two particles moving with a relative angle larger than
$1/\gamma$; if they move in
the same direction their distance may not increase, but
the density in their center of mass frame is diluted by 
a Lorentz factor and becomes too low 
to give interactions \cite{MacGibbon:2007yq} 
($\theta<1/\gamma$ defines an {\it exclusion cone}). 
The key observation is that
each quark can complete at most one bremsstrahlung
interaction before crossing $r_{brem}$, so after that
interaction its (tranverse) distance with the 
particles out of the exclusion cone
will be much larger than $1/\Lambda_{QCD}$. 
To understand that, let
us suppose that, right when it is created next to the 
BH horizon, a quark $q$ absorbs a virtual gluon of 
$Q^2\approx \Lambda^2$ and goes off shell. 
In the $q$ rest frame both its energy
and its off-shellness after absorbing the gluon are
of order $\Lambda$, which in a QCD bremsstrahlung should 
be just above $\Lambda_{QCD}$. Then the virtual $q$ 
lives a time of order $1/\Lambda$ and 
decays into the final quark and gluon. Now, going 
back to the BH frame we observe the lifetime of the virtual 
gluon a Lorentz factor larger, so the typical
distance that it travels will be of order 
$\gamma /\Lambda \gg 1/\Lambda_{QCD} \approx r_{brem}$.
Except for small values of $\gamma$ ({\it i.e.}, non-relativistic 
quarks emitted by BHs of temperature close to $\Lambda_{QCD}$)
this argument establishes that quarks exiting the BH
cannot interact with each other more than once, as would
be necessary to form a chromosphere. Our conclusion agrees
with the simulation of BH production and evaporation at
the LHC in \cite{Alig:2006up}.

\section{Spectrum of stable particles}
The cosmogenic BHs under study here are produced at 
astrophysical distances 
from the Earth, therefore, unstable particles resulting
from their evaporation have plenty of time to decay.
A BH emits stable neutrinos, 
electrons, photons and gravitons, but mostly, it emits quarks 
and gluons that will fragment into hadrons and then shower 
into stable particles. In this section we evaluate the total
spectrum of stable species from a BH of mass up to
$10^{11}$ GeV. Our results are analogous to the ones obtained
by MacGibbon and Webber in \cite{MacGibbon:1990zk} for primordial 
BHs \cite{Carr:2005zd}. 
Of course, we use an updated MonteCarlo jet code
(HERWIG 6.5 \cite{Corcella:2000bw}) and include the effects of the 
extra dimensions,
namely, appropriate greybody factors and bulk graviton emission. 
In addition, while in primordial BHs the spectrum corresponds
to a given temperature ($T$ changes only within 
astrophysical time scales), here we evaluate the total spectrum
resulting from the complete evaporation of the BH, which includes
a (relatively small) contribution from the high
temperatures briefly reached at the end of its lifetime. 

\begin{figure}
\begin{center}
\begin{tabular}{cc}
\includegraphics[width=0.5\linewidth]{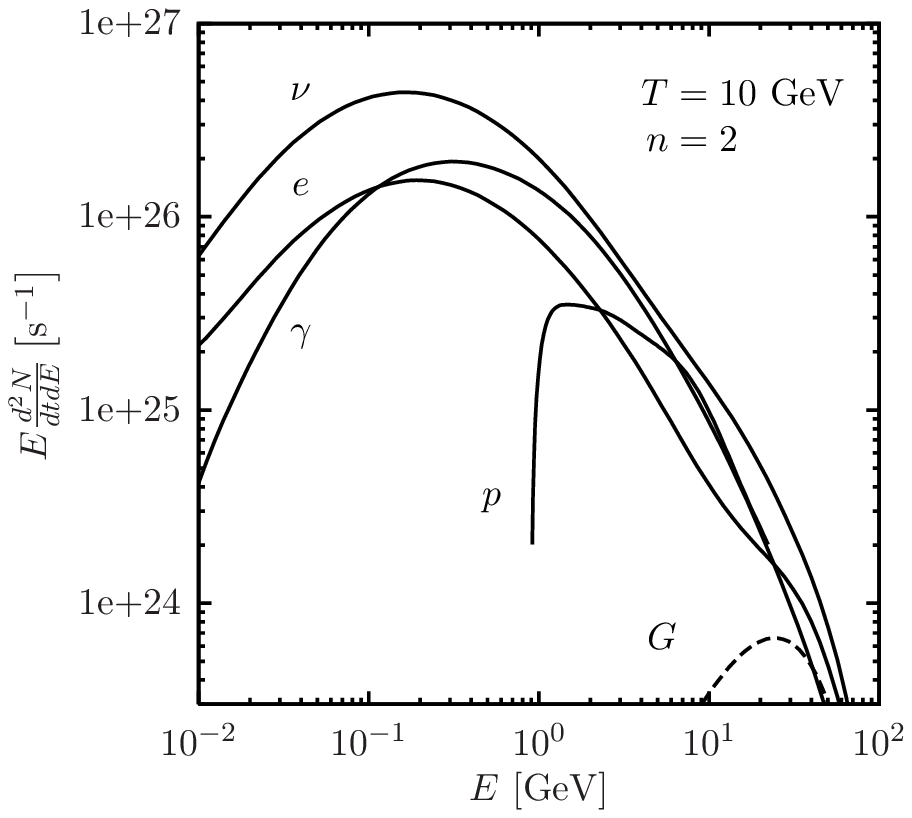} &
\includegraphics[width=0.5\linewidth]{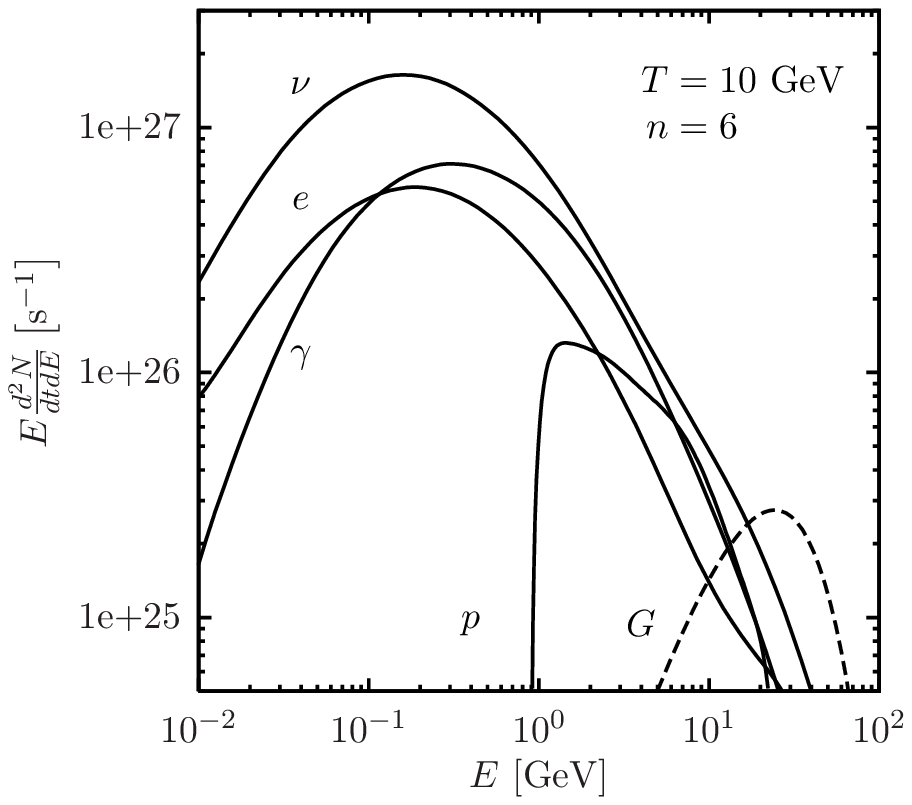} 
\end{tabular}
\end{center}
\caption{Spectrum of stable particles and bulk gravitons (dashed)
at $T=10$ GeV for $M_D=1$ TeV, $n=2$ (left) and $n=6$ (right).
\label{fig5}}
\end{figure}

Several comments are here in order. As we have explained in the 
previous section, when the quarks and gluons exit the BH 
their probability of interaction is
small. We will then assume that the 
jets that they define are similar
to the ones produced in $e^+ e^-$ collisions 
to $q\overline q$ or (the fictitious process) $gg$.
We have used HERWIG \cite{Corcella:2000bw} to
simulate the jets produced by any particle that is light 
at a given temperature, including 
massive gauge bosons and the top quark (but not the Higgs
boson nor a dark matter particle). 
Finally, we have added together the number of 
particles and antiparticles (they are produced at the
same rate) and the three (Majorana) neutrino species 
(at astrophysical distances their flavour oscillates).

In Fig.~5 we plot the power spectrum emitted by a BH of 
$T=10$ GeV for $n=2,6$. We observe that it is
dominated by energies around $\Lambda_{QCD}$, although 
the primary greybody spectrum peaks at 30 GeV. This is
manifest in the flux of gravitons, since they {\it decouple}
(their number is not increased by the showering of unstable 
particles). At high energies 
it is possible to distinguish the primary greybody spectra
for some of the particle species. 
At this temperature the
particles emitted onto the brane consist of an approximate 
43\% of neutrinos, a 28\% of photons, a 16\% of electrons and 
a 13\% of protons. The bulk gravitons are
a 0.02\% of the total emitted particles for $n=2$ or a 0.4\% for
$n=6$. Their typical energy is higher, so they
account for a 1.4\% ($n=2$) or a 16\% ($n=6$) of the total 
energy radiated. 

\begin{figure}
\begin{center}
\begin{tabular}{cc}
\includegraphics[width=0.5\linewidth]{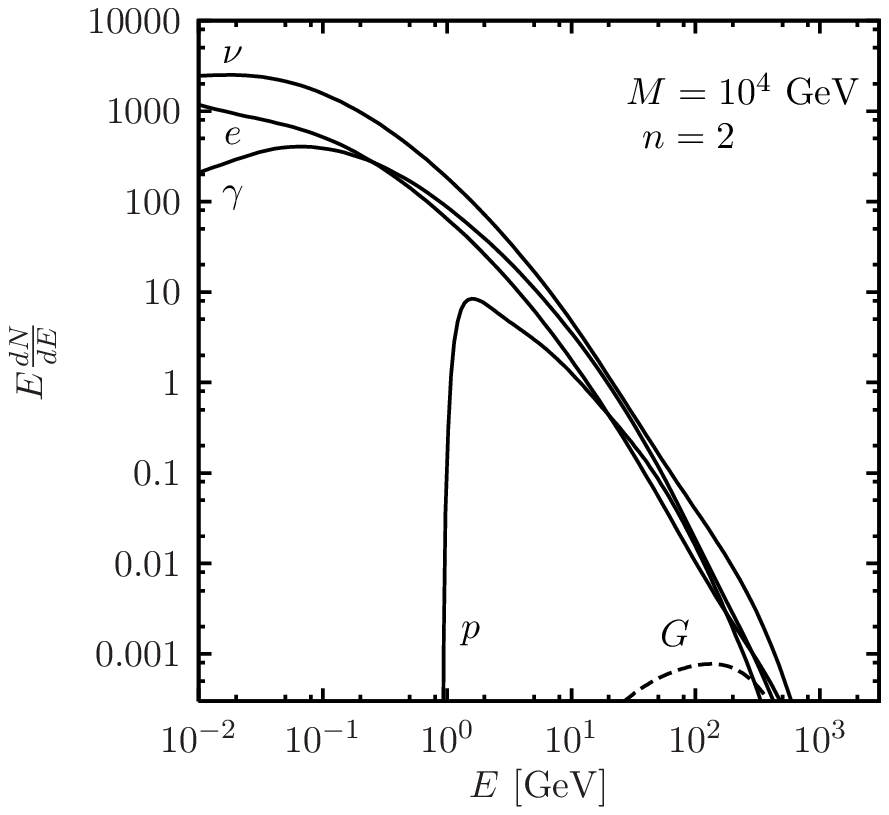} &
\includegraphics[width=0.5\linewidth]{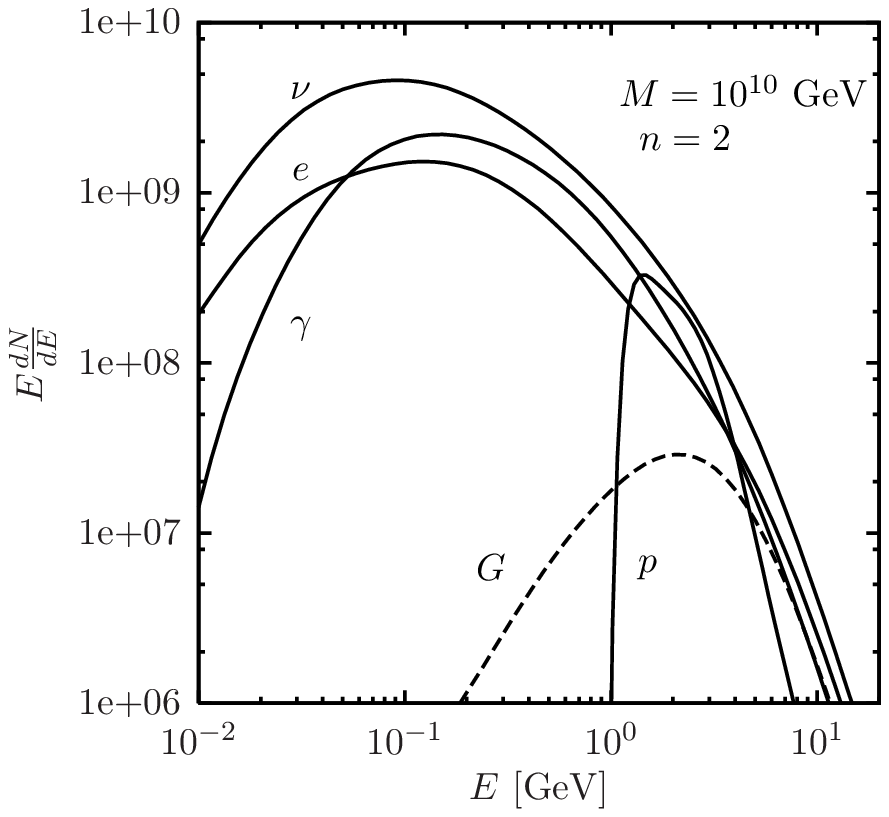} 
\end{tabular}
\end{center}
\caption{Spectrum of stable particles and bulk gravitons (dashed)
from a BH of $M=10$ TeV (left) and $M=10^{10}$ GeV for $M_D=1$ TeV 
and $n=2$.
\label{fig6}}
\end{figure}

In Fig.~6 we give the total spectrum from BHs of mass
$10$ TeV and $10^{10}$ GeV for $n=2$. These masses correspond 
to initial temperatures of 120 and 1.2 GeV, respectively. 
We find that the spectra are dominated by the emission at 
these initial temperatures, in particular, 90\% of the energy
is emitted by the $10^{10}$ GeV BH when its temperature has 
only increased from 1.2 to 2.6 GeV (see Fig.~\ref{fig3}). 
The dominance in the spectrum of energies around 
0.2 GeV and the approximate fraction of different species are features 
that depend very mildly on the BH mass, as it is apparent
in these plots.
Of course, in {\it smaller} BHs of higher initial temperature 
the relative weight of high energies in the spectrum is larger.

\section{Summary and discussion}
Models with the scale of gravity at the TeV must
confront the fact that in nature there are processes
at much higher energy. In particular, in the collision of
two cosmic rays the center of mass energy can go up to 
$10^{11}$ GeV. At these energies for impact parameters
short enough the particles will be trapped inside 
a gravitational horizon and form a mini BH. We have
estimated the production rate of these 
cosmogenic BHs (in Fig.~2)
and have analyzed their properties. 

{\it (i)} We have found  that their lifetime goes from 
$10^{-26}$ s for a light BH of $M=10$ TeV to $10^{-14}$ s 
for $M=10^{11}$ GeV. In Fig.~\ref{fig3} we plot the 
correlation between BH mass, temperature and lifetime for
$n=2,6$. 

{\it (ii)} We have shown that bremsstrahlung interactions
between the particles exiting the BH are unable to form a 
chromosphere. Although the average separation between these 
particles (in Table 1) can be in some cases small, a 
QCD process requires a typical time to develop, and 
this time is such that the particle can not interact 
more than once at $r < 1/\Lambda_{QCD}$ in the BH frame.

{\it (iii)} The greybody spectrum emitted by the BH 
onto the brane is
dominated by colored particles. These form jets that
result into a spectrum of stable particles peaked at 
$\Lambda_{QCD}$. We obtain an approximate 43\% of 
neutrinos, a 28\% of photons, a 16\% of electrons and 
a 13\% of protons. These two features in the spectrum
depend very 
mildly on the number of extra dimensions (Fig.~5)
or the BH mass (Fig.~6).

{\it (iv)} Bulk graviton emission is relatively larger 
for high BH masses and large number of extra dimensions,
and it is peaked at higher energies than for the rest of
species. It accounts for a 0.4\% of the total particles 
(16\% of the energy) emitted by a $10^{10}$ GeV BH
for $n=6$ or just a 0.02\% (1.4\% of the energy) 
for $M=10$ TeV and $n=2$.

We find remarkable that, even though there is no chromosphere,
the spectrum of stable particles from the evaporation is 
equally peaked at low (0.2 GeV) energies: the spectrum provided 
by one of these BHs at astrophysical distances would not be
too different whether there is or there is not a chromosphere 
around it, in both cases the scale is fixed by $\Lambda_{QCD}$.
Notice, however, that the signal at the LHC in one case or 
the other would be clearly different (10 jets of 100 GeV each
versus an expanding shell of 1000 hadrons).

We think that the analysis presented here is a necessary 
first step in the 
search for possible observable effects from BH production 
by cosmic rays.

\section*{Acknowledgements}
This work has been supported by MEC of Spain 
(FPA2006-05294), by
Junta de Andaluc\'\i a (FQM-101 and FQM-437),
and by the European Community's Marie-Curie Research
Training Network under contract MRTN-CT-2006-035505 {\it Tools
and Precision Calculations for Physics Discoveries
at Colliders}. I.M. acknowledges a grant from 
C.F. Luciano Fonda (Italy).


\begin{thebibliography}{99}

\bibitem{ArkaniHamed:1998rs}
  N.~Arkani-Hamed, S.~Dimopoulos and G.~R.~Dvali,
  Phys.\ Lett.\  B {\bf 429} (1998) 263;
  I.~Antoniadis, N.~Arkani-Hamed, S.~Dimopoulos and G.~R.~Dvali,
  Phys.\ Lett.\  B {\bf 436} (1998) 257;
  L.~Randall and R.~Sundrum,
  Phys.\ Rev.\ Lett.\  {\bf 83} (1999) 3370.

\bibitem{Yao:2006px}
  W.~M.~Yao {\it et al.}  [Particle Data Group],
  J.\ Phys.\ G {\bf 33} (2006) 1.

\bibitem{Banks:1999gd}
  T.~Banks and W.~Fischler,
  ``A model for high energy scattering in quantum gravity,''
  arXiv:hep-th/9906038;
  R.~Emparan,
  Phys.\ Rev.\  D {\bf 64} (2001) 024025;
  S.~B.~Giddings and S.~D.~Thomas,
  Phys.\ Rev.\  D {\bf 65} (2002) 056010.
  D.~M.~Eardley and S.~B.~Giddings,
  Phys.\ Rev.\  D {\bf 66} (2002) 044011.

\bibitem{Cullen:2000ef}
  S.~Cullen, M.~Perelstein and M.~E.~Peskin,
  Phys.\ Rev.\  D {\bf 62} (2000) 055012;
  F.~Cornet, J.~I.~Illana and M.~Masip,
  Phys.\ Rev.\ Lett.\  {\bf 86} (2001) 4235.

\bibitem{Giudice:2001ce}
  G.~F.~Giudice, R.~Rattazzi and J.~D.~Wells,
  Nucl.\ Phys.\  B {\bf 630} (2002) 293.

\bibitem{Cheung:2002aq}
  S.~Dimopoulos and G.~L.~Landsberg,
  Phys.\ Rev.\ Lett.\  {\bf 87} (2001) 161602;
  K.~Cheung,
  Phys.\ Rev.\  D {\bf 66} (2002) 036007;
  T.~G.~Rizzo,
  JHEP {\bf 0202}, 011 (2002);
  A.~Chamblin and G.~C.~Nayak,
  Phys.\ Rev.\  D {\bf 66}, 091901 (2002);
  R.~Casadio and B.~Harms,
  Int.\ J.\ Mod.\ Phys.\  A {\bf 17} (2002) 4635;
  M.~Cavaglia,
  Phys.\ Lett.\  B {\bf 569} (2003) 7;
  I.~Mocioiu, Y.~Nara and I.~Sarcevic,
  Phys.\ Lett.\  B {\bf 557}, 87 (2003);
  M.~Cavaglia, S.~Das and R.~Maartens,
  Class.\ Quant.\ Grav.\  {\bf 20}, L205 (2003);
  A.~Barrau, J.~Grain and S.~O.~Alexeyev,
  Phys.\ Lett.\  B {\bf 584}, 114 (2004);
  C.~M.~Harris, M.~J.~Palmer, M.~A.~Parker, P.~Richardson, 
  A.~Sabetfakhri and B.~R.~Webber,
  JHEP {\bf 0505}, 053 (2005);
  J.~Tanaka, T.~Yamamura, S.~Asai and J.~Kanzaki,
  Eur.\ Phys.\ J.\  C {\bf 41}, 19 (2005);
  J.~L.~Hewett, B.~Lillie and T.~G.~Rizzo,
  Phys.\ Rev.\ Lett.\  {\bf 95}, 261603 (2005);
  B.~Koch, M.~Bleicher and S.~Hossenfelder,
  JHEP {\bf 0510}, 053 (2005);
  D.~M.~Gingrich,
  Int.\ J.\ Mod.\ Phys.\  A {\bf 21} (2006) 6653;
  T.~J.~Humanic, B.~Koch and H.~Stoecker,
  Int.\ J.\ Mod.\ Phys.\  E {\bf 16}, 841 (2007);
  G.~L.~Alberghi, R.~Casadio and A.~Tronconi,
  J.\ Phys.\ G {\bf 34} (2007) 767;
  T.~G.~Rizzo,
  Phys.\ Lett.\  B {\bf 647}, 43 (2007);
  M.~Cavaglia, R.~Godang, L.~M.~Cremaldi and D.~J.~Summers,
  JHEP {\bf 0706}, 055 (2007);
  G.~Dvali and S.~Sibiryakov,
  JHEP {\bf 0803} (2008) 007.

\bibitem{Harris:2003db}
  C.~M.~Harris, P.~Richardson and B.~R.~Webber,
  JHEP {\bf 0308} (2003) 033;
  M.~Cavaglia, R.~Godang, L.~Cremaldi and D.~Summers,
  Comput.\ Phys.\ Commun.\  {\bf 177}, 506 (2007);
  D.~C.~Dai, G.~Starkman, D.~Stojkovic, C.~Issever, E.~Rizvi and J.~Tseng,
  Phys.\ Rev.\  D {\bf 77} (2008) 076007.

\bibitem{Winstanley:2007hj}
  E.~Winstanley,
  arXiv:0708.2656 [hep-th].

\bibitem{Meade:2007sz}
  P.~Meade and L.~Randall,
  arXiv:0708.3017 [hep-ph].

\bibitem{Cavaglia:2002si}
  M.~Cavaglia,
  Int.\ J.\ Mod.\ Phys.\  A {\bf 18} (2003) 1843

\bibitem{Corcella:2000bw}
  G.~Corcella {\it et al.},
  JHEP {\bf 0101} (2001) 010.

\bibitem{Feng:2001ib}
  J.~L.~Feng and A.~D.~Shapere,
  Phys.\ Rev.\ Lett.\  {\bf 88} (2002) 021303;
  L.~Anchordoqui and H.~Goldberg,
  Phys.\ Rev.\  D {\bf 65} (2002) 047502;
  R.~Emparan, M.~Masip and R.~Rattazzi,
  Phys.\ Rev.\  D {\bf 65} (2002) 064023.
  L.~A.~Anchordoqui, J.~L.~Feng, H.~Goldberg and A.~D.~Shapere,
  Phys.\ Rev.\  D {\bf 65} (2002) 124027;
  A.~Ringwald and H.~Tu,
  Phys.\ Lett.\  B {\bf 525} (2002) 135;
  M.~Kowalski, A.~Ringwald and H.~Tu,
  Phys.\ Lett.\  B {\bf 529} (2002) 1;
  S.~I.~Dutta, M.~H.~Reno and I.~Sarcevic,
  Phys.\ Rev.\  D {\bf 66} (2002) 033002;
  J.~Alvarez-Muniz, J.~L.~Feng, F.~Halzen, T.~Han and D.~Hooper,
  Phys.\ Rev.\  D {\bf 65} (2002) 124015;
  L.~A.~Anchordoqui, J.~L.~Feng, H.~Goldberg and A.~D.~Shapere,
  Phys.\ Rev.\  D {\bf 68} (2003) 104025;
  J.~I.~Illana, M.~Masip and D.~Meloni,
  Phys.\ Rev.\ Lett.\  {\bf 93} (2004) 151102;
  E.~J.~Ahn, M.~Cavaglia and A.~V.~Olinto,
  Astropart.\ Phys.\  {\bf 22} (2005) 377;
  J.~I.~Illana, M.~Masip and D.~Meloni,
  Phys.\ Rev.\  D {\bf 72} (2005) 024003.

\bibitem{Barrau:2005zb}
  A.~Barrau, C.~Feron and J.~Grain,
  Astrophys.\ J.\  {\bf 630} (2005) 1015.

\bibitem{Semikoz:2003wv}
  D.~V.~Semikoz and G.~Sigl,
  JCAP {\bf 0404} (2004) 003.

\bibitem{Feng:2008ya}
  J.~L.~Feng and J.~Kumar,
  arXiv:0803.4196 [hep-ph].

\bibitem{Navarro:1995iw}
  J.~F.~Navarro, C.~S.~Frenk and S.~D.~M.~White,
  Astrophys.\ J.\  {\bf 462} (1996) 563.

\bibitem{Hawking:1974rv}
  S.~W.~Hawking,
  Nature {\bf 248} (1974) 30;
  S.~W.~Hawking,
  Commun.\ Math.\ Phys.\  {\bf 43} (1975) 199
  [Erratum-ibid.\  {\bf 46} (1976) 206].

\bibitem{Emparan:2000rs}
  R.~Emparan, G.~T.~Horowitz and R.~C.~Myers,
  Phys.\ Rev.\ Lett.\  {\bf 85} (2000) 499
  [arXiv:hep-th/0003118].

\bibitem{Page:1976df}
  D.~N.~Page,
  Phys.\ Rev.\  D {\bf 13} (1976) 198;
  Phys.\ Rev.\  D {\bf 14} (1976) 3260;
  Phys.\ Rev.\  D {\bf 16} (1977) 2402.

\bibitem{Harris:2003eg}
  C.~M.~Harris and P.~Kanti,
  JHEP {\bf 0310} (2003) 014;
  P.~Kanti and J.~March-Russell,
  Phys.\ Rev.\  D {\bf 66} (2002) 024023;
  P.~Kanti and J.~March-Russell,
  Phys.\ Rev.\  D {\bf 67} (2003) 104019.

\bibitem{Cardoso:2005vb}
  V.~Cardoso, M.~Cavaglia and L.~Gualtieri,
  Phys.\ Rev.\ Lett.\  {\bf 96} (2006) 071301
  [Erratum-ibid.\  {\bf 96} (2006) 219902].

\bibitem{Heckler:1995qq}
  A.~F.~Heckler,
  Phys.\ Rev.\  D {\bf 55} (1997) 480.

\bibitem{Cline:1998xk}
  J.~M.~Cline, M.~Mostoslavsky and G.~Servant,
  Phys.\ Rev.\  D {\bf 59} (1999) 063009;
  R.~G.~Daghigh and J.~I.~Kapusta,
  Phys.\ Rev.\  D {\bf 65} (2002) 064028;
  A.~Casanova and E.~Spallucci,
  Class.\ Quant.\ Grav.\  {\bf 23} (2006) R45.

\bibitem{Anchordoqui:2002cp}
  L.~Anchordoqui and H.~Goldberg,
  Phys.\ Rev.\  D {\bf 67} (2003) 064010
  [arXiv:hep-ph/0209337].

\bibitem{Alig:2006up}
  C.~Alig, M.~Drees and K.~y.~Oda,
  JHEP {\bf 0612} (2006) 049.

\bibitem{MacGibbon:2007yq}
  J.~H.~MacGibbon, B.~J.~Carr and D.~N.~Page,
  ``Do Evaporating Black Holes Form Photospheres?,''
  arXiv:0709.2380 [astro-ph];
  D.~N.~Page, B.~J.~Carr and J.~H.~MacGibbon,
  ``Bremsstrahlung Effects around Evaporating Black Holes,''
  arXiv:0709.2381 [astro-ph].

\bibitem{Klein:1998du}
  S.~Klein,
  Rev.\ Mod.\ Phys.\  {\bf 71} (1999) 1501.

\bibitem{MacGibbon:1990zk}
  J.~H.~MacGibbon and B.~R.~Webber,
  Phys.\ Rev.\  D {\bf 41} (1990) 3052.

\bibitem{Carr:2005zd}
  B.~J.~Carr,
  arXiv:astro-ph/0511743.









 
\end{thebibliography}
\end{document}